\documentclass[aps,prl,twocolumn,superscriptaddress,showpacs,preprintnumbers,amsmath,amssymb]{revtex4}
\usepackage{graphicx} 
\usepackage{dcolumn}  

\graphicspath{{ps}}

\begin{document}


\preprint{\vbox{ \hbox{   }
                 \hbox{BELLE-CONF-0713}
}}

\title{ \quad\\[0.5cm]  Search for lepton flavor violating 
$\tau \to \ell V^0$ decays at Belle}



\affiliation{Budker Institute of Nuclear Physics, Novosibirsk}
\affiliation{Chiba University, Chiba}
\affiliation{University of Cincinnati, Cincinnati, Ohio 45221}
\affiliation{Department of Physics, Fu Jen Catholic University, Taipei}
\affiliation{Justus-Liebig-Universit\"at Gie\ss{}en, Gie\ss{}en}
\affiliation{The Graduate University for Advanced Studies, Hayama}
\affiliation{Gyeongsang National University, Chinju}
\affiliation{Hanyang University, Seoul}
\affiliation{University of Hawaii, Honolulu, Hawaii 96822}
\affiliation{High Energy Accelerator Research Organization (KEK), Tsukuba}
\affiliation{Hiroshima Institute of Technology, Hiroshima}
\affiliation{University of Illinois at Urbana-Champaign, Urbana, Illinois 61801}
\affiliation{Institute of High Energy Physics, Chinese Academy of Sciences, Beijing}
\affiliation{Institute of High Energy Physics, Vienna}
\affiliation{Institute of High Energy Physics, Protvino}
\affiliation{Institute for Theoretical and Experimental Physics, Moscow}
\affiliation{J. Stefan Institute, Ljubljana}
\affiliation{Kanagawa University, Yokohama}
\affiliation{Korea University, Seoul}
\affiliation{Kyoto University, Kyoto}
\affiliation{Kyungpook National University, Taegu}
\affiliation{\'Ecole Polytechnique F\'ed\'erale de Lausanne (EPFL), Lausanne}
\affiliation{University of Ljubljana, Ljubljana}
\affiliation{University of Maribor, Maribor}
\affiliation{University of Melbourne, School of Physics, Victoria 3010}
\affiliation{Nagoya University, Nagoya}
\affiliation{Nara Women's University, Nara}
\affiliation{National Central University, Chung-li}
\affiliation{National United University, Miao Li}
\affiliation{Department of Physics, National Taiwan University, Taipei}
\affiliation{H. Niewodniczanski Institute of Nuclear Physics, Krakow}
\affiliation{Nippon Dental University, Niigata}
\affiliation{Niigata University, Niigata}
\affiliation{University of Nova Gorica, Nova Gorica}
\affiliation{Osaka City University, Osaka}
\affiliation{Osaka University, Osaka}
\affiliation{Panjab University, Chandigarh}
\affiliation{Peking University, Beijing}
\affiliation{University of Pittsburgh, Pittsburgh, Pennsylvania 15260}
\affiliation{Princeton University, Princeton, New Jersey 08544}
\affiliation{RIKEN BNL Research Center, Upton, New York 11973}
\affiliation{Saga University, Saga}
\affiliation{University of Science and Technology of China, Hefei}
\affiliation{Seoul National University, Seoul}
\affiliation{Shinshu University, Nagano}
\affiliation{Sungkyunkwan University, Suwon}
\affiliation{University of Sydney, Sydney, New South Wales}
\affiliation{Tata Institute of Fundamental Research, Mumbai}
\affiliation{Toho University, Funabashi}
\affiliation{Tohoku Gakuin University, Tagajo}
\affiliation{Tohoku University, Sendai}
\affiliation{Department of Physics, University of Tokyo, Tokyo}
\affiliation{Tokyo Institute of Technology, Tokyo}
\affiliation{Tokyo Metropolitan University, Tokyo}
\affiliation{Tokyo University of Agriculture and Technology, Tokyo}
\affiliation{Toyama National College of Maritime Technology, Toyama}
\affiliation{Virginia Polytechnic Institute and State University, Blacksburg, Virginia 24061}
\affiliation{Yonsei University, Seoul}
  \author{K.~Abe}\affiliation{High Energy Accelerator Research Organization (KEK), Tsukuba} 
  \author{I.~Adachi}\affiliation{High Energy Accelerator Research Organization (KEK), Tsukuba} 
  \author{H.~Aihara}\affiliation{Department of Physics, University of Tokyo, Tokyo} 
  \author{K.~Arinstein}\affiliation{Budker Institute of Nuclear Physics, Novosibirsk} 
  \author{T.~Aso}\affiliation{Toyama National College of Maritime Technology, Toyama} 
  \author{V.~Aulchenko}\affiliation{Budker Institute of Nuclear Physics, Novosibirsk} 
  \author{T.~Aushev}\affiliation{\'Ecole Polytechnique F\'ed\'erale de Lausanne (EPFL), Lausanne}\affiliation{Institute for Theoretical and Experimental Physics, Moscow} 
  \author{T.~Aziz}\affiliation{Tata Institute of Fundamental Research, Mumbai} 
  \author{S.~Bahinipati}\affiliation{University of Cincinnati, Cincinnati, Ohio 45221} 
  \author{A.~M.~Bakich}\affiliation{University of Sydney, Sydney, New South Wales} 
% ...
\collaboration{The Belle Collaboration}

\begin{abstract}
We have searched for neutrinoless $\tau$ lepton decays into $\ell$ and
$V^0$, where $\ell$ stands for an electron or muon, and 
$V^0$ for a vector meson ($\phi$, $\omega$, $K^{*0}$ or $\bar{K}^{*0}$), 
using 543~fb$^{-1}$ of data collected
with the Belle detector at the KEKB asymmetric-energy $e^+e^-$ collider.
No excess of signal events over the expected
background  is observed, and we set upper limits on the 
branching fractions in the range $(0.7 - 1.8) \times 10^{-7}$ at the 
90\% confidence level. These upper limits 
include the first results for $\ell \omega$ as well as new limits that are
$3.6 - 9.6$ times more restrictive than our previous results for $\ell \phi$,
$\ell K^{*0}$ and $\ell \bar{K}^{*0}$.
\end{abstract}


\maketitle

\tighten

{\renewcommand{\thefootnote}{\fnsymbol{footnote}}}
\setcounter{footnote}{0}

\section{Introduction}

In the Standard Model (SM), lepton-flavor-violating (LFV) decays of charged
leptons are forbidden; 
even if neutrino mixing is taken into account, they are still highly suppressed. 
However, LFV is expected to appear in many extensions of the SM. 
Some such models predict branching fractions
for $\tau$ LFV decays at the level of $10^{-8}-10^{-7}$~\cite{Amon,BR,CG},
which can be reached at the present B-factories.
Observation of LFV  will then provide evidence for new physics 
beyond the SM.

In this paper, we report on a search for LFV in $\tau^-$ decays
into neutrinoless final states with one charged lepton $\ell^-$ and
one vector meson $V^0$: $e^- \phi$, $e^- \omega$, $e^- K^{*0}$,
$e^- \bar{K}^{*0}$, $\mu^- \phi$, $\mu^- \omega$, $\mu^- K^{*0}$ and
$\mu^- \bar{K}^{*0}$~\cite{CC}. 
A search for the $\ell^- \phi$, $\ell^- K^{*0}$ and $\ell^- \bar{K}^{*0}$ 
modes was performed for the first time at the CLEO detector, where
90\% confidence level (CL)
upper limits (UL) for the branching fractions in the range 
$(5.1 - 7.5) \times 10^{-6}$ were obtained using a data sample of 
4.79 fb$^{-1}$~\cite{CLEO}.
Later we carried out a search for these modes in the Belle experiment 
using 158~fb$^{-1}$ of data and set
upper limits in the range $(3.0 - 7.7) \times 10^{-7}$~\cite{previous_Belle}.
Here we present results of a new search based on a data sample of 
543~fb$^{-1}$ corresponding to $4.99 \times 10^8$ $\tau$-pairs 
collected with the Belle
detector~\cite{Belle} at the KEKB asymmetric-energy $e^+ e^-$ 
collider~\cite{KEKB}.

The Belle detector is a large-solid-angle magnetic spectrometer that
consists of a silicon vertex detector,
a 50-layer central drift chamber, an array of
aerogel threshold Cherenkov counters, 
a barrel-like arrangement of time-of-flight
scintillation counters, and an electromagnetic calorimeter
comprised of CsI(Tl) crystals located inside 
a superconducting solenoid coil that provides a 1.5~T
magnetic field.  An iron flux-return located outside 
the coil is instrumented to detect $K_L^0$ mesons and identify
muons.  The detector is described in detail elsewhere~\cite{Belle}.
Two inner detector configurations were used. A 2.0 cm radius beam-pipe
and a 3-layer silicon vertex detector were used for the first sample
of 158~fb$^{-1}$, while a 1.5 cm radius beam-pipe, a 4-layer
silicon detector and a small-cell inner drift chamber were used to record  
the remaining 385~fb$^{-1}$~\cite{svd2}.  

\section{Event Selection}

We search for $\tau \to \ell \phi$, $\ell \omega$, $\ell K^{*0}$
and $\ell \bar{K}^{*0}$ candidates in 
which one $\tau$ decays into a final state with a $\ell$,
two charged hadrons (3-prong decay), and the other $\tau$ decays 
into one charged particle (1-prong decay), any number of $\gamma$'s
 and missing particle(s).
We reconstruct $\phi$ candidates from $K^+ K^-$,
$\omega$ from $\pi^+\pi^-\pi^0$,
$K^{*0}$ from $K^+ \pi^-$ and $\bar{K}^{*0}$ from $K^- \pi^+$.

The selection criteria described below are optimized 
from studies of Monte Carlo (MC) simulated events
and the experimental data in the sideband regions
of the $\Delta E$ and $M_{\rm inv}$ distributions described later.
The background (BG) MC samples consist of $\tau^+\tau^-$ (1524~fb$^{-1}$)
generated by KKMC~\cite{KKMC},
$q\overline{q}$ continuum, and two-photon processes. 
The signal MC events are generated assuming a phase space distribution
for $\tau$ decay.

The transverse momentum for a charged track is required to be larger than 
0.06 GeV/$c$ in the barrel region ($-0.6235<\cos \theta<0.8332$, where 
$\theta$ is the polar angle relative to the direction opposite to that of
the incident $e^+$ beam in the laboratory frame)
and 0.1 GeV/$c$ in the endcap region ($-0.8660<\cos \theta<-0.6235$ and 
$0.8332<\cos \theta<0.9563$).
The energies of photon candidates are required to be 
larger than 0.1 GeV in both regions. 

To select the signal topology,
we require four charged tracks in an event with zero net charge,
and a total energy of charged tracks and photons
in the center-of-mass (CM) frame less than 11 GeV.
We also require that the missing momentum in the laboratory frame be 
greater than 0.6 GeV/$c$, and that its direction be within the
detector acceptance ($-0.8660 < \cos \theta < 0.9563$),
where the missing momentum is defined as the 
difference between the momentum of the initial $e^+e^-$
system, and the sum of the observed momentum vectors.
The event is subdivided into 3-prong and 1-prong
hemispheres with respect to the thrust axis in the CM frame.
These are referred to as the signal and tag side, respectively.
We allow at most two photons on the tag side
to account for initial state radiation,
while requiring at most one photon for the $\ell \phi$, 
$\ell K^{*0}$, $\ell \bar{K}^{*0}$ modes,
and two photons except for $\pi^0$ daughters for the $\ell \omega$ 
modes on the signal side to reduce the $q\bar{q}$ BG. 

We require that
the muon likelihood ratio $P_{\mu}$ be greater than 0.95
for momentum greater than 1.0 GeV/$c$
and
the electron likelihood ratio $P_e$ be greater than 0.9
for momentum greater than 0.5 GeV/$c$
for the charged lepton-candidate track on the signal side.
Here $P_x$ is the likelihood ratio for a charged particle
of type $x$ ($x = \mu$, $e$, $K$ or $\pi$), defined as 
$P_x = L_x/(\sum_x L_x)$, 
where $L_x$ is the likelihood for particle type $x$,
determined from the responses of the relevant detectors~\cite{LID}.
The efficiencies for muon and electron identification are
92\% for momenta larger than 1.0 GeV/$c$ and
94\% for momenta larger than 0.5 GeV/$c$.

\begin{figure}[htb]
\includegraphics[height=0.23\textwidth]{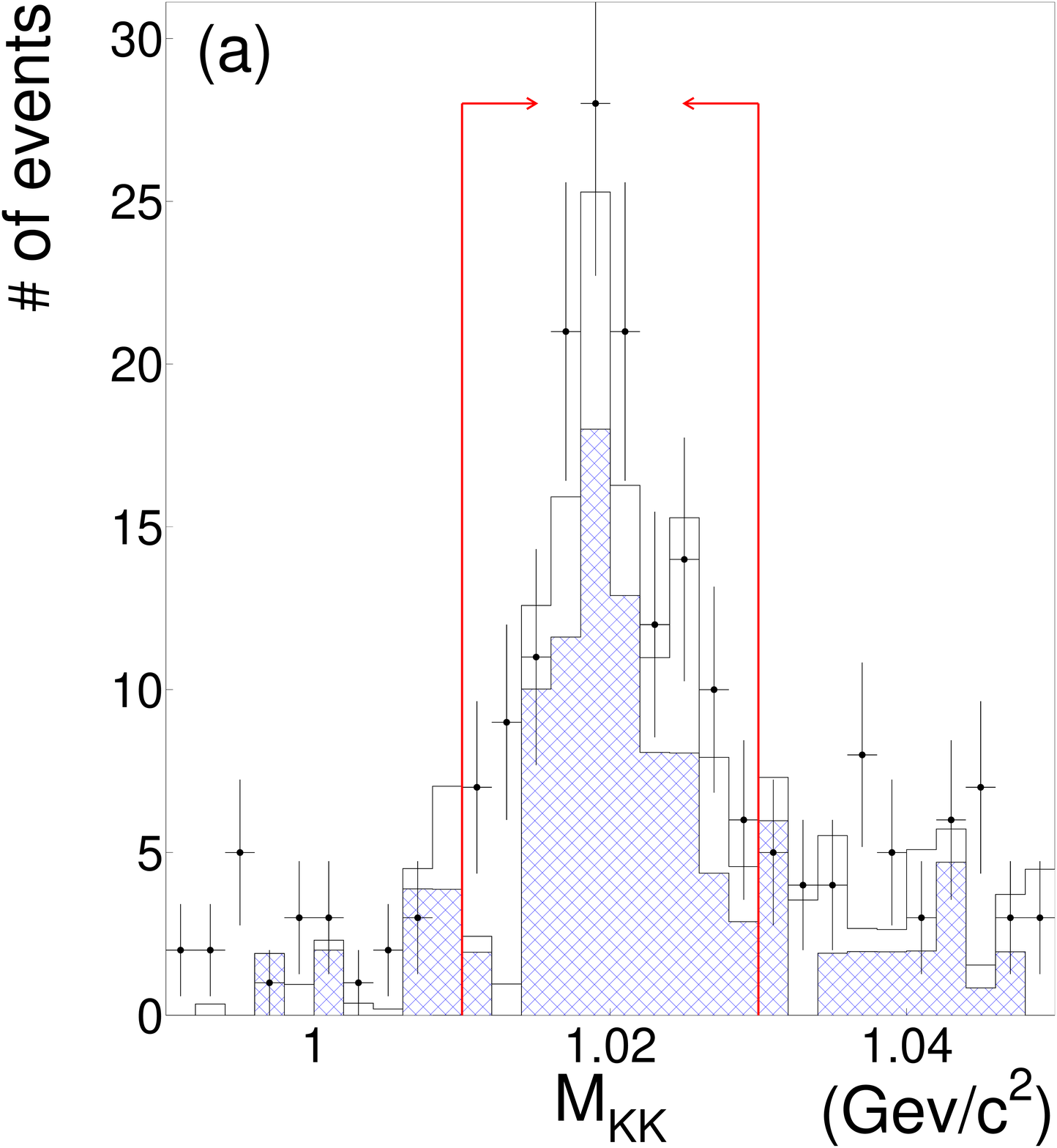}
\includegraphics[height=0.23\textwidth]{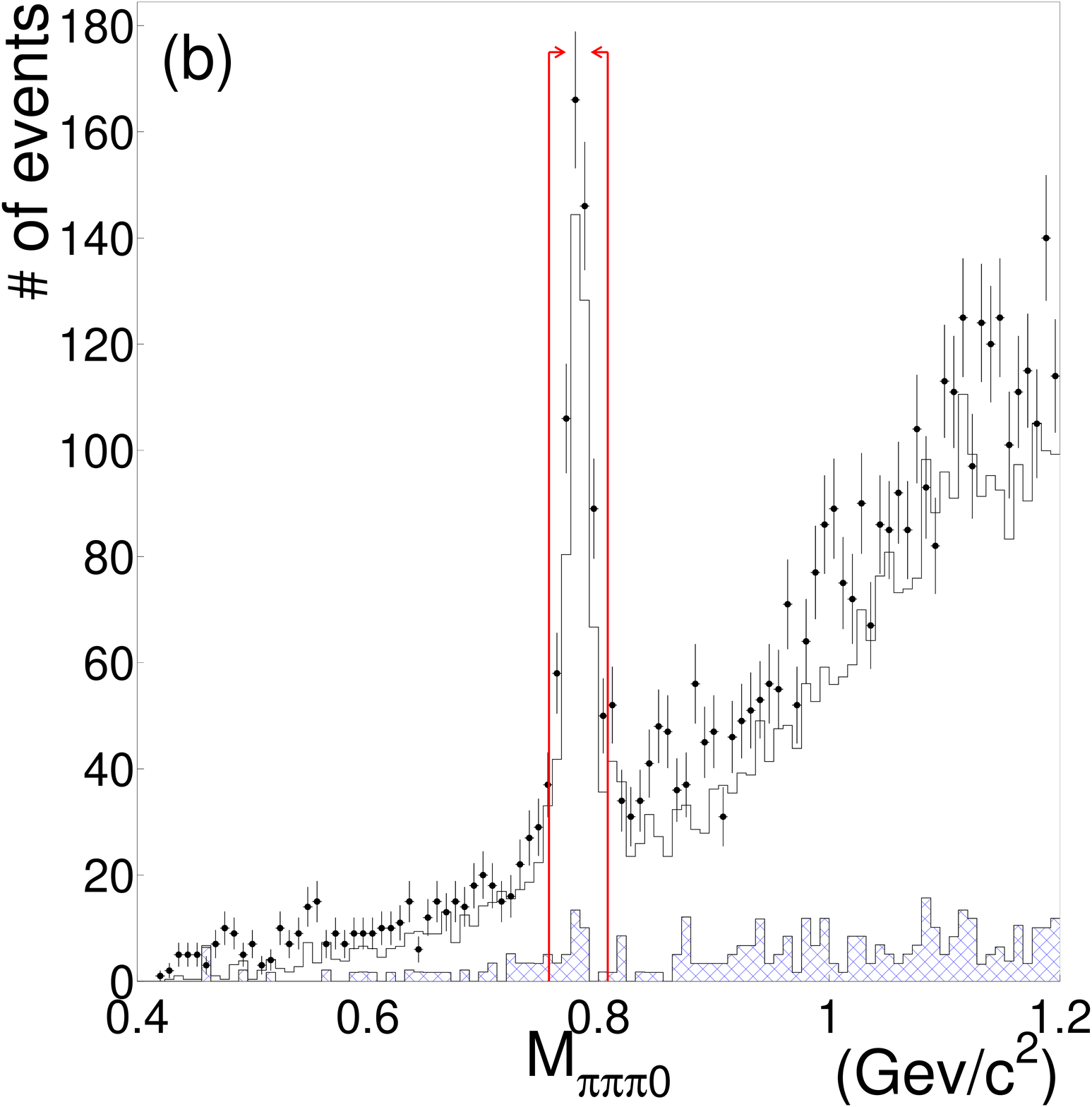}
\includegraphics[height=0.23\textwidth]{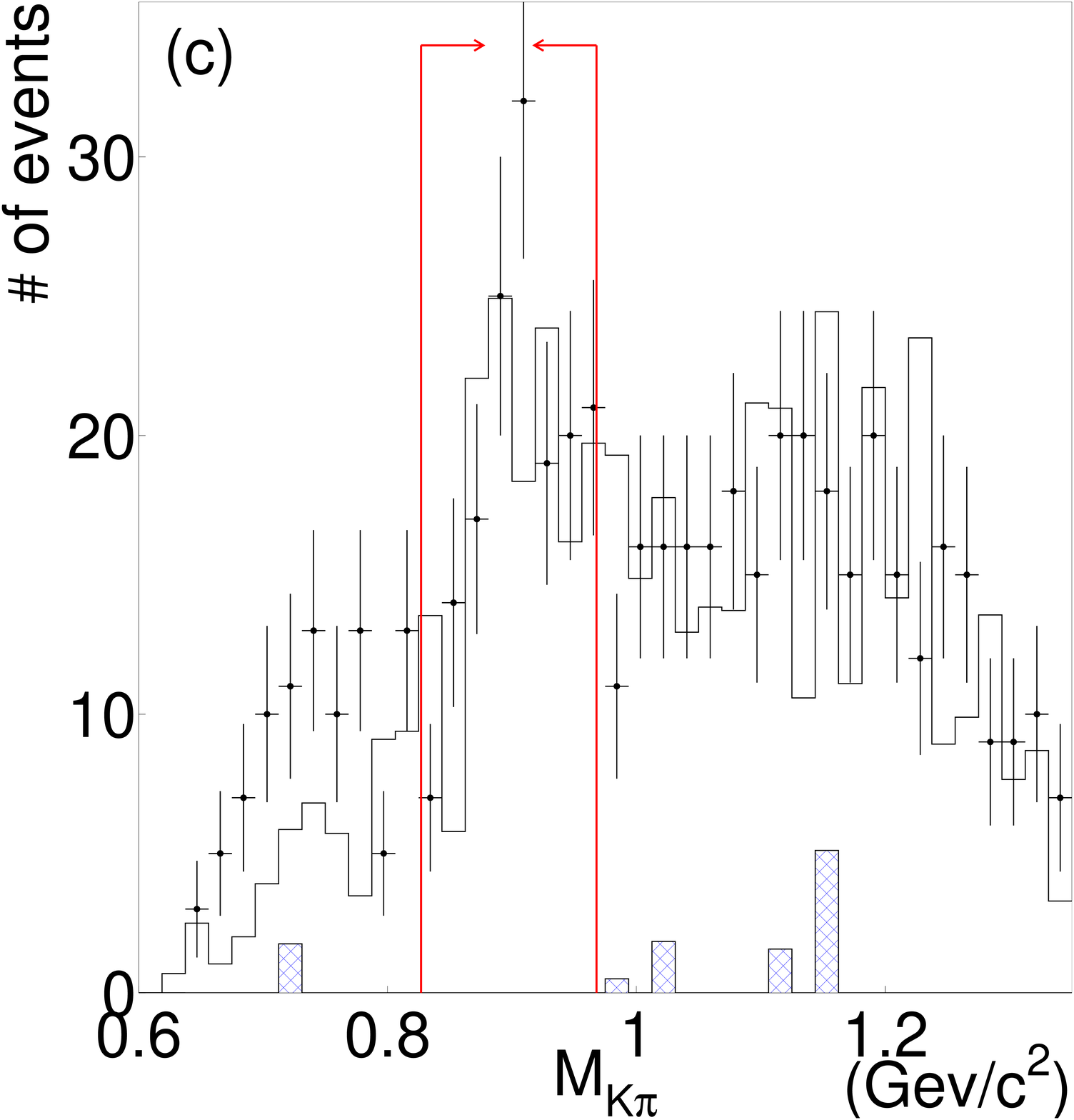}
\caption{The mass distribution of (a) $\phi \to K^+ K^-$ for
$\tau^- \to \mu^- \phi$, (b) $\omega \to \pi^+ \pi^- \pi^0$ for
$\tau^- \to \mu^- \omega$ and
(c) $K^{*0} \to K^+ \pi^-$ for $\tau^- \to \mu^- K^{*0}$
after muon identification.
The points with error bars are data.
The open histogram shows the expected $\tau^+\tau^-$ BG MC
and the hatched one $q\bar{q}$ MC and two-photon MC.
The regions between the vertical red lines are selected.}
\label{fig:Vmass}
\end{figure}

Candidate $\phi$ mesons are selected by requiring the invariant mass of
$K^+ K^-$ daughters to be in the range 
$1.01~{\rm GeV}/c^2~<~M_{K^+K^-}~<~1.03~{\rm GeV}/c^2$.
We require that both kaon daughters have kaon likelihood ratios $P_K > 0.8$
and electron likelihood ratios $P_e < 0.1$ 
to reduce the background from $e^+ e^-$ conversions.
Candidate $\omega$ mesons are reconstructed from $\pi^+\pi^-\pi^0$
with the invariant mass requirement  
$0.757~{\rm GeV}/c^2~<~M_{\pi^+\pi^-\pi^0}~<~0.808~{\rm GeV}/c^2$.
The $\pi^0$ candidate is selected from $\gamma$ pairs 
with invariant mass in the range,
$0.11~{\rm GeV}/c^2~<~M_{\gamma \gamma}~<~0.15~{\rm GeV}/c^2$.
In order to improve the $\omega$ mass resolution,
the $\pi^0$ mass is constrained to be 135 MeV/$c^2$ for
the $\omega$ mass reconstruction.
Candidate $K^{*0}$ and $\bar{K}^{*0}$ mesons are selected with 
$K^{\pm}\pi^{\mp}$ 
invariant mass in the range 
$0.827~{\rm GeV}/c^2~<~M_{K\pi}~<~0.986~{\rm GeV}/c^2$,
and requiring that the kaon daughter have $P_K > 0.8$
and both daughters have $P_e < 0.1$.
Figs.~\ref{fig:Vmass}(a,b,c) show the invariant mass distributions of
the $\phi$, $\omega$ and $K^{*0}$ candidates for $\tau^- \to \mu^- \phi$,
$\tau^- \to \mu^- \omega$ and $\tau^- \to \mu^- K^{*0}$, respectively.
The estimated BG distributions agree with the data.
The main BG contribution is due to $q\bar{q}$ events with $\phi$
mesons for the $\tau^- \to \ell^- \phi$ mode,
 $\tau^- \to \pi^- \omega \nu_{\tau}$ with the pion misidentified as a lepton 
for the $\tau^- \to \ell^- \omega$ mode,
and 
$\tau^- \to \pi^- \pi^+ \pi^- \nu_{\tau}$ with one pion 
misidentified as a kaon and another misidentified as
a lepton for the $\tau^- \to \ell^- K^{*0}$ and $\ell^- \bar{K}^{*0}$ modes.

To reduce the remaining BG from $\tau^+\tau^-$ and $q\bar{q}$,
we require the relations between 
the missing momentum $p_{\rm miss}$ (GeV/$c$)
and missing mass squared $m^2_{\rm miss}$ ((GeV/$c^2$)$^2$) summarized 
in Table~\ref{tbl:vcut}.
\begin{table}
\caption{Selection criteria using $p_{\rm miss}$(GeV/$c$) 
and $m^2_{\rm miss}$((GeV/$c^2$)$^2$) where $p_{\rm miss}$ is 
 missing momentum and $m^2_{\rm miss}$ is missing mass squared.}
\label{tbl:vcut}
\begin{center}
\begin{tabular}{c|c} \hline
Mode &  Selection criteria  \\ \hline
$\tau^- \to \ell^- \phi$ & $p_{\rm miss} > \frac{8}{9} m^2_{\rm miss}$ and $m^2_{\rm miss}> -0.5$ \\ 
$\tau^- \to \ell^- \omega$ & $p_{\rm miss} > \frac{8}{3} m^2_{\rm miss} -\frac{8}{3} $ and $m^2_{\rm miss}> -0.5$  \\ 
 $\tau^-\to \mu^- K^{*0}$  & $p_{\rm miss} > \frac{8}{4.5} m^2_{\rm miss}-\frac{8}{9}$ and $p_{\rm miss} > 8 m^2_{\rm miss}$ \\
 $\tau^-\to e^- K^{*0}$  & $p_{\rm miss} > \frac{8}{5.5} m^2_{\rm miss}-\frac{8}{11}$ and $m^2_{\rm miss}>0$ \\
$\tau^- \to \mu^- \bar{K}^{*0}$  & $p_{\rm miss} > \frac{8}{6.5} m^2_{\rm miss}$ and $m^2_{\rm miss}>-0.5$ \\
$\tau^- \to e^- \bar{K}^{*0}$  & $p_{\rm miss} > \frac{6}{5} m^2_{\rm miss}$ and $p_{\rm miss} > -\frac{8}{1.4} m^2_{\rm miss}$ \\ \hline
\end{tabular}
\end{center}
\end{table}

For the $\ell \omega$ ($\ell K^{*0}$ and $\ell \bar{K}^{*0}$) mode, 
we require that the opening angle between 
the lepton and $\omega$ ($K^{*0}$) on the signal side in the CM frame,
$\theta_{\ell \omega}^{\rm CM}$ ($\theta_{\ell K^{*0}}^{\rm CM}$), satisfy 
$\cos \theta_{\ell \omega}^{\rm CM} < 0.88$
($\cos \theta_{\ell K^{*0}}^{\rm CM} < 0.93$), respectively.
To remove two-photon BG for the $e \phi$, $e \omega$
and $e K^{*0} (\bar{K}^{*0})$ modes,
we add a condition on an opening angle, $\alpha$, between the direction of
the total momentum of charged tracks and $\gamma$'s on the signal side
and that on the tag side, as $\cos \alpha > -0.999$,
$\cos \alpha > -0.996$ and
$\cos \alpha > -0.990$, respectively.

To identify signal $\tau$ decays,
we reconstruct the invariant mass of $\ell V^0$, $M_{\rm inv}$, and
the energy difference, $\Delta E$, between the sum of energies on 
the signal side and the beam energy $E_{\rm beam}$ in the CM frame.
Signal events should be distributed around $M_{\rm inv} = M_\tau$ and
$\Delta E = 0$, where $M_\tau$ is the nominal $\tau$ mass.
For the $\ell \omega$ modes, we use the beam energy constrained mass 
$M_{\rm bc}$ as the invariant mass $M_{\rm inv}$,
where $M_{\rm bc} = \sqrt{E_{\rm beam}^2 - (\vec{p}_\tau)^2}$,
in order to improve the mass resolution, which is smeared 
due to the $\gamma$ energy resolution.
For the calculation of the $\tau$ momentum $\vec{p}_\tau$,
we replace the magnitude of the $\pi^0$ momentum 
with the momentum calculated from the beam energy and 
the energies of charged tracks on the signal side,
while we fix the direction of the $\pi^0$ momentum.

The resolutions in $\Delta E$ and $M_{\rm inv}$ 
evaluated using the signal MC are summarized in Table~\ref{tbl:res}.
We define the signal region in the $\Delta E - M_{\rm inv}$ plane
as a $\pm 3 \sigma$ ellipse.
In order to avoid biases in the event selection,
we blinded the signal region until the analysis is finalized
and used the data in a $\pm 10 \sigma$ sideband box
to estimate BG.

\begin{table}[htb]
\caption{Resolutions in $M_{\rm inv}$ in MeV/$c^2$ and $\Delta E$ in MeV.
The superscripts low and high indicate the lower and higher sides of 
the peak, respectively.}
\label{tbl:res}
\begin{center}
\begin{tabular}{ccccc}
\hline
Mode & ~~~$\sigma_{M_{\rm inv}}^{\rm high}$~~~ & ~~~$\sigma_{M_{\rm inv}}^{\rm low}$~~~ & ~~~~$\sigma_{\Delta E}^{\rm high}$~~~~ & ~~~~$\sigma_{\Delta E}^{\rm low}$~~~~ \\
\hline
$\tau^-\to\mu^-\phi$   & $3.4\pm0.2$ & $3.4\pm0.2$ & $13.2\pm0.4$ & $14.0\pm0.5$ \\
$\tau^-\to e^- \phi$   & $3.7\pm0.1$ & $3.6\pm0.1$ & $13.3\pm0.7$ & $15.4\pm0.7$ \\
$\tau^-\to\mu^-\omega$ & $5.9\pm0.1$ & $6.2\pm0.1$ & $19.3\pm0.6$ & $30.3\pm0.8$ \\
$\tau^-\to e^- \omega$ & $6.1\pm0.1$ & $6.5\pm0.1$ & $20.4\pm0.7$ & $32.5\pm1.3$ \\
$\tau^-\to\mu^- K^{*0}$& $4.5\pm0.4$ & $4.5\pm0.4$ & $13.8\pm0.3$ & $14.4\pm0.4$ \\
$\tau^-\to e^- K^{*0}$ & $4.3\pm0.1$ & $5.1\pm0.1$ & $12.9\pm0.3$ & $18.0\pm0.4$ \\
$\tau^-\to\mu^- \bar{K}^{*0}$ & $4.7\pm0.1 $ & $4.4\pm0.1$ & $14.0\pm0.3$ & $15.0\pm0.3$ \\  
$\tau^-\to e^-\bar{K}^{*0}$ & $4.6\pm0.1$ & $4.9\pm0.1$ & $12.6\pm0.6$ & $17.8\pm0.5$ \\
\hline
\end{tabular}
\end{center}
\end{table}

\section{Results}

\begin{figure}[htb]
\includegraphics[width=0.23\textwidth]{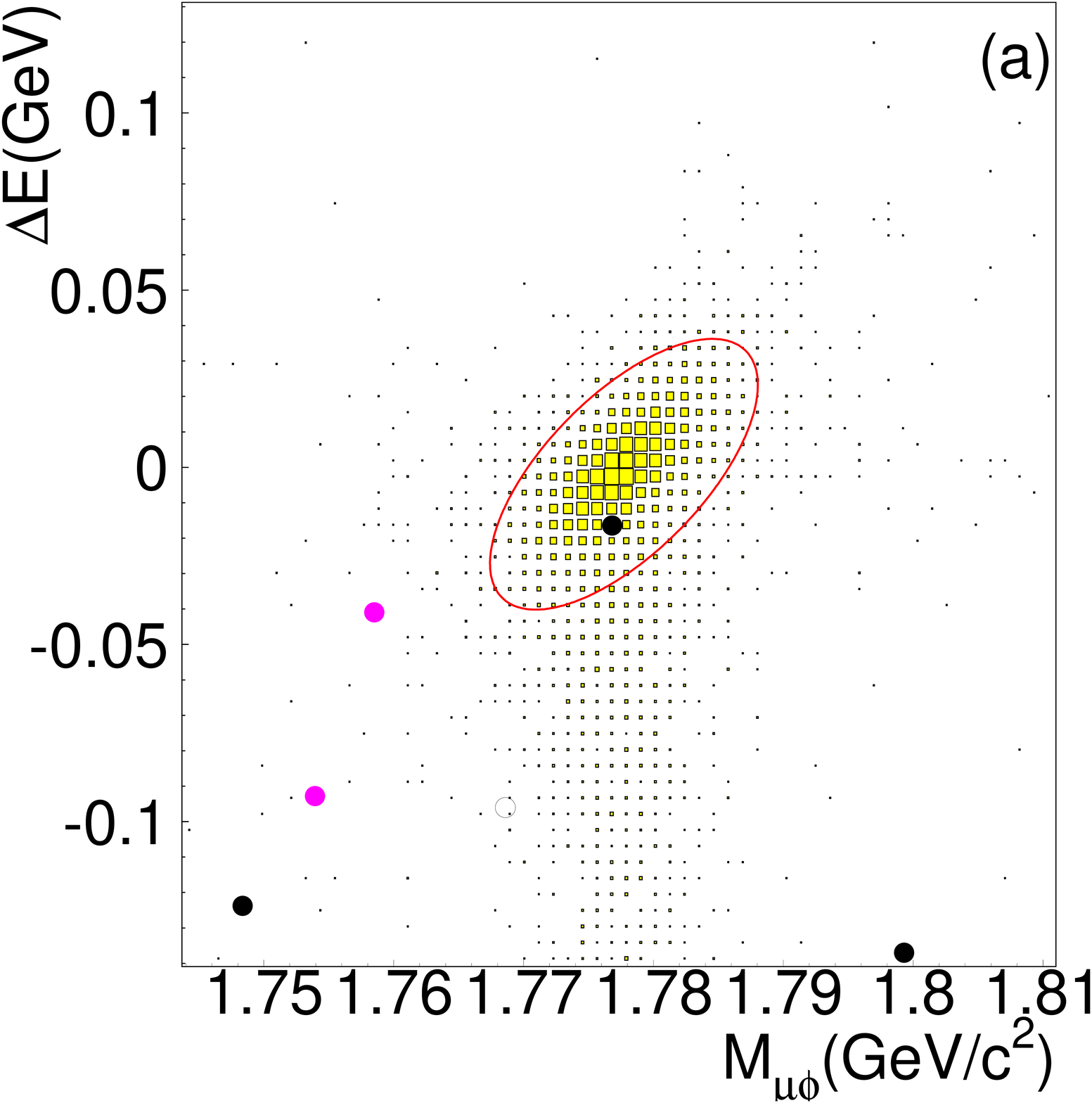}
\includegraphics[width=0.23\textwidth]{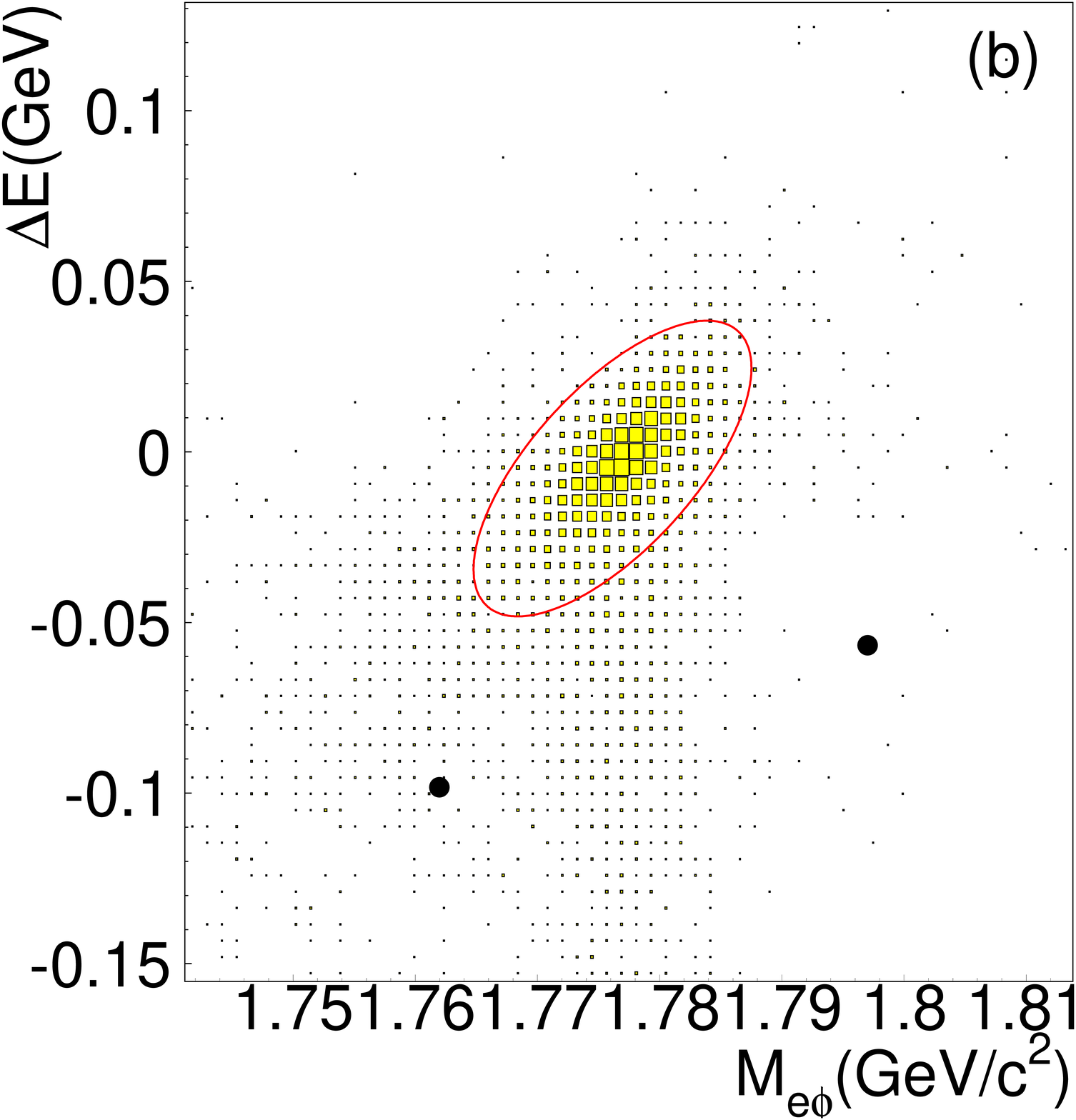}
\caption{$\Delta E$ vs. $M_{\rm inv}$ distributions for 
(a) $\tau^- \to \mu^- \phi$ and (b) $\tau^- \to e^- \phi$, after all selection
criteria. Dots are data, yellow boxes show the signal MC, purple dots are 
background from $\tau \to \phi \pi \nu$ MC and open circles show other
$\tau$-pair backgrounds. The elliptical area is the $3\sigma$ signal region.}
\label{fig:result1}
\end{figure}

\begin{figure}[htb]
\includegraphics[width=0.23\textwidth]{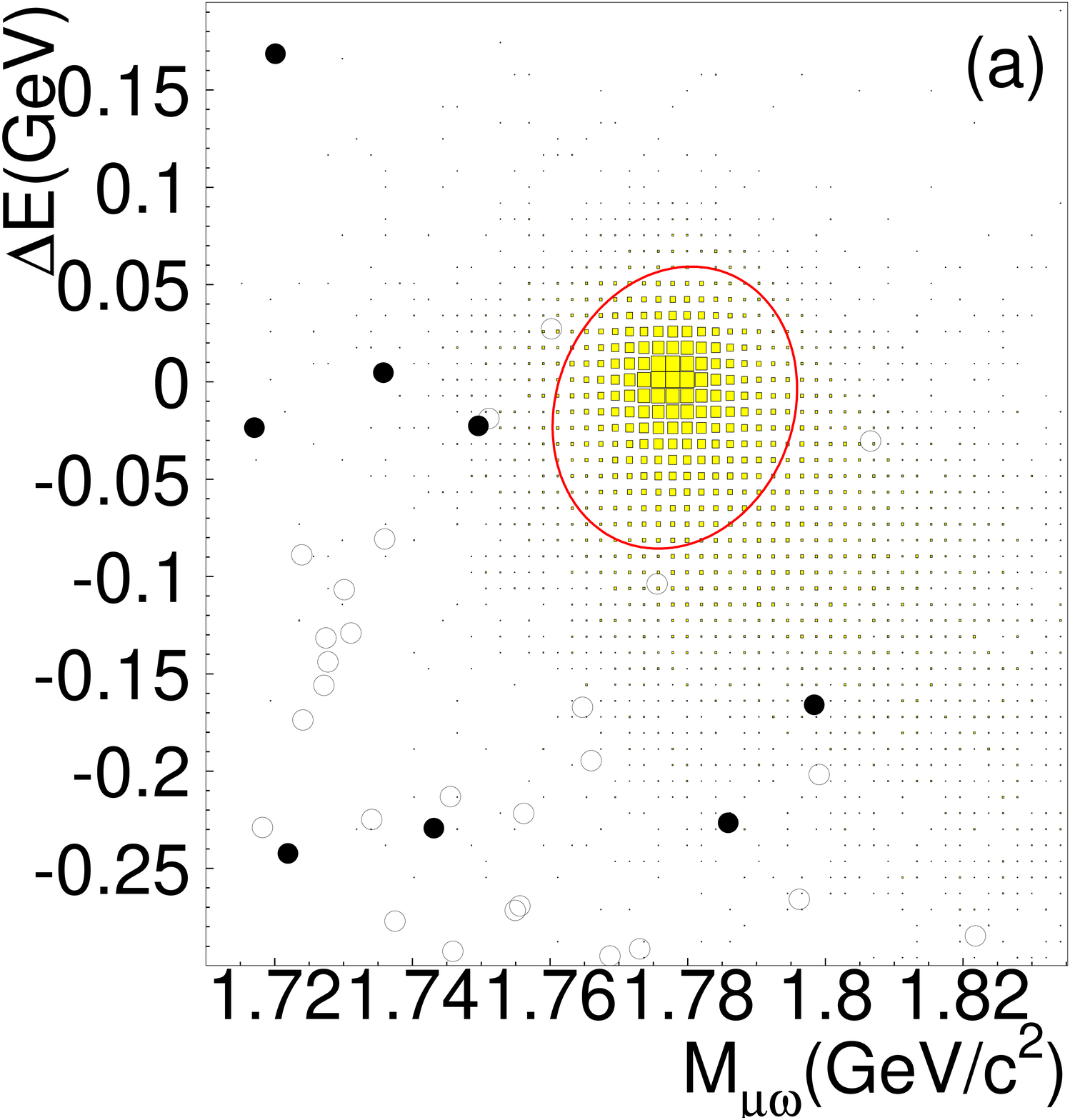}
\includegraphics[width=0.23\textwidth]{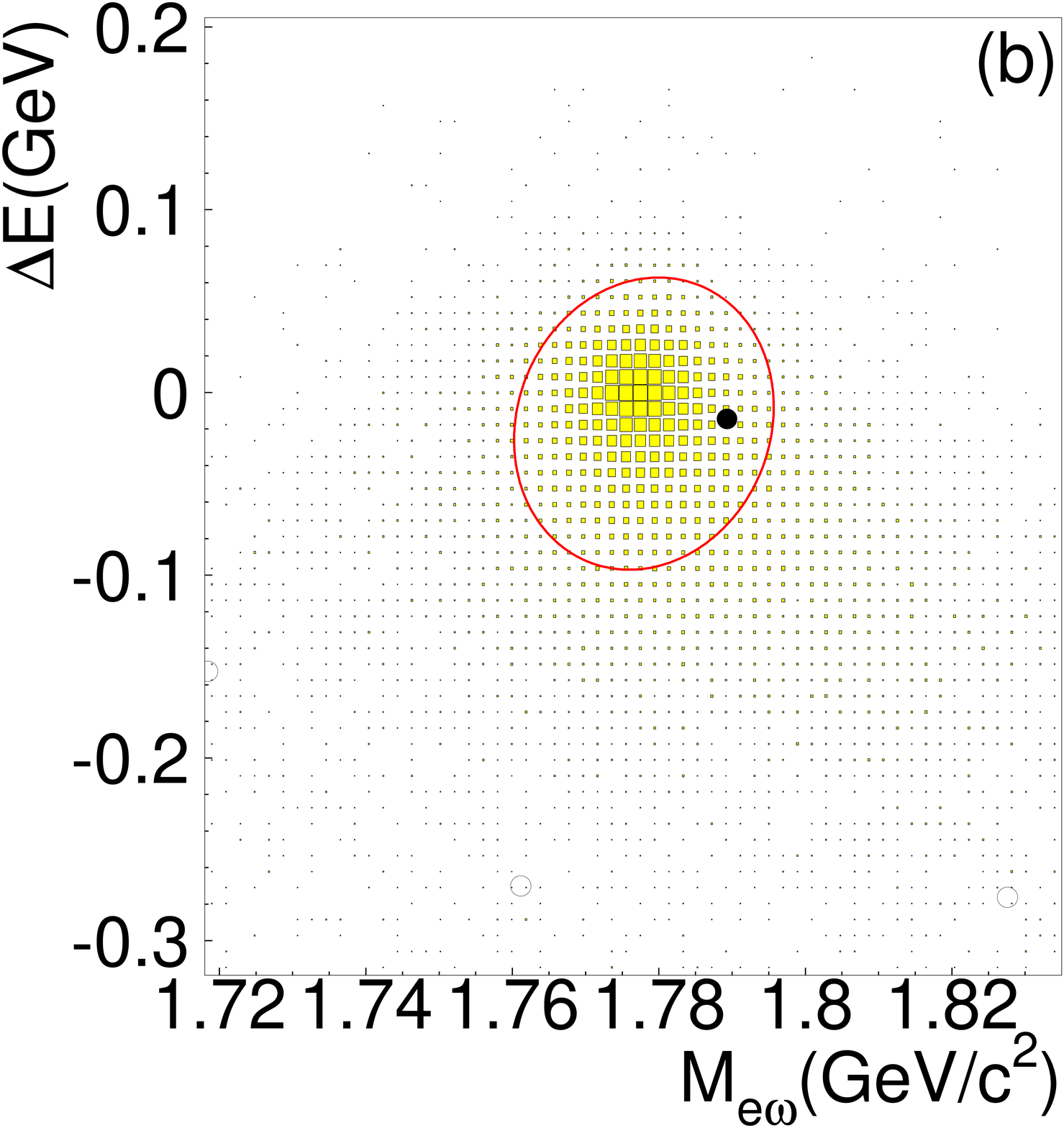}
\caption{$\Delta E$ vs. $M_{\rm inv}$ distributions for 
(a) $\tau^- \to \mu^- \omega$ and (b) $\tau^- \to e^- \omega$, 
after all selection criteria. Dots are data, yellow boxes show the signal MC, 
and open circles show
$\tau$-pair background MC. The elliptical area is the $3\sigma$ signal region.}
\label{fig:result2}
\end{figure}

\begin{figure}[htb]
\includegraphics[width=0.23\textwidth]{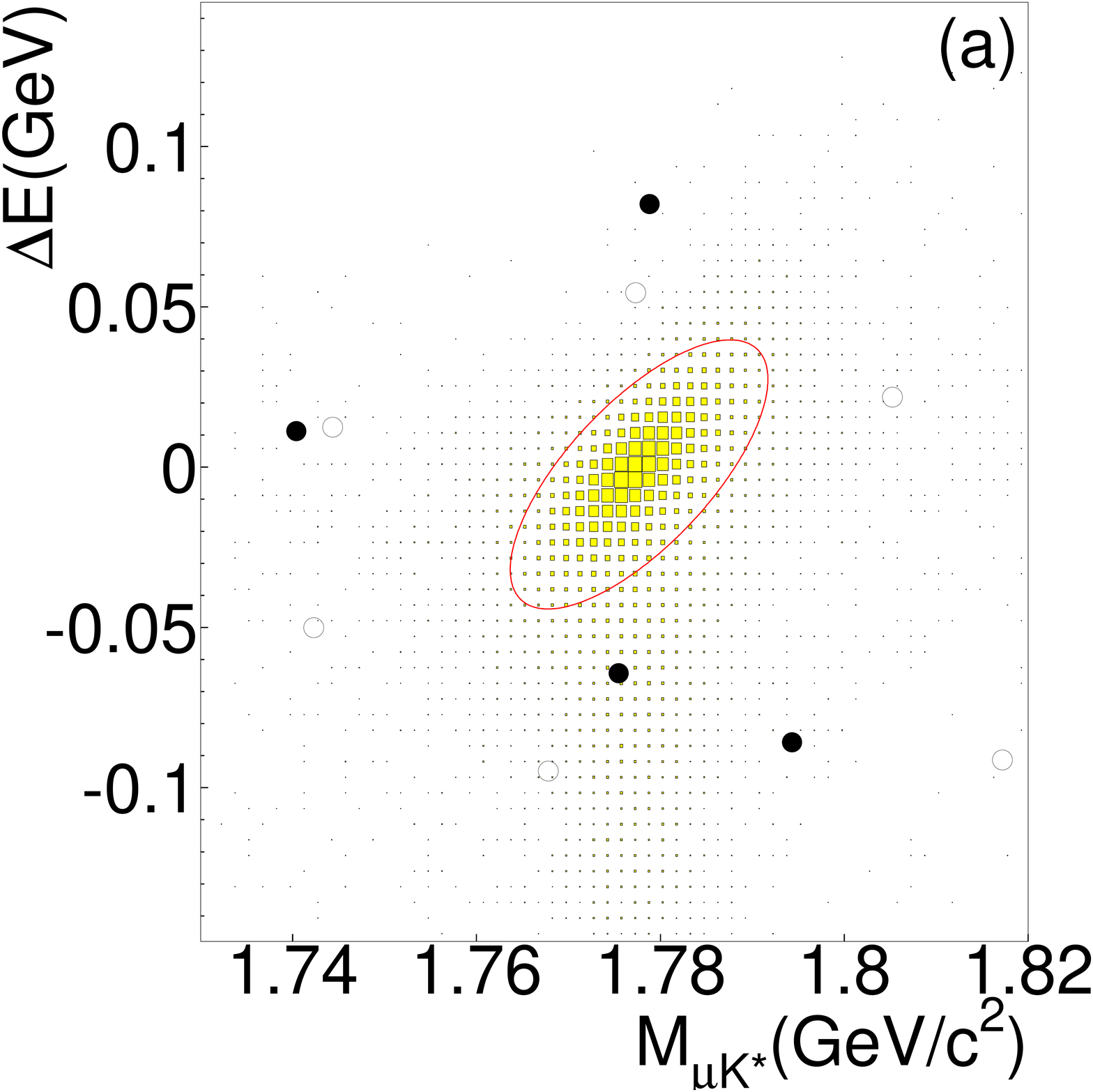}
\includegraphics[width=0.23\textwidth]{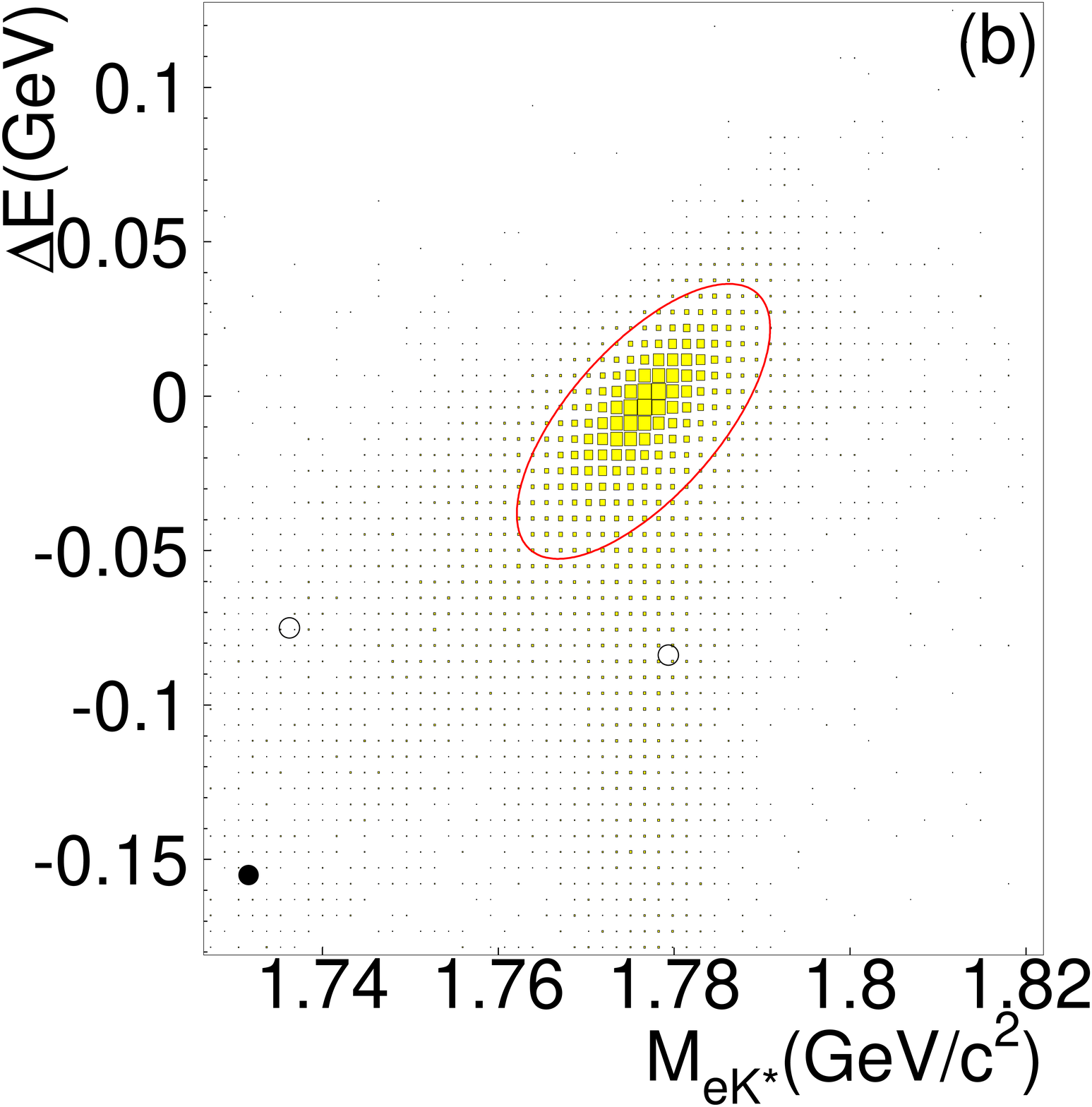}
\caption{$\Delta E$ vs. $M_{\rm inv}$ distributions for 
(a) $\tau^- \to \mu^- K^{*0}$ and (b) $\tau^- \to e^- K^{*0}$,
after all selection criteria.
Dots are data, yellow boxes show the signal MC and open circles show
$\tau$-pair background MC. The elliptical area is the $3\sigma$ signal region.}
\label{fig:result3}
\end{figure}

\begin{figure}[htb]
\includegraphics[width=0.23\textwidth]{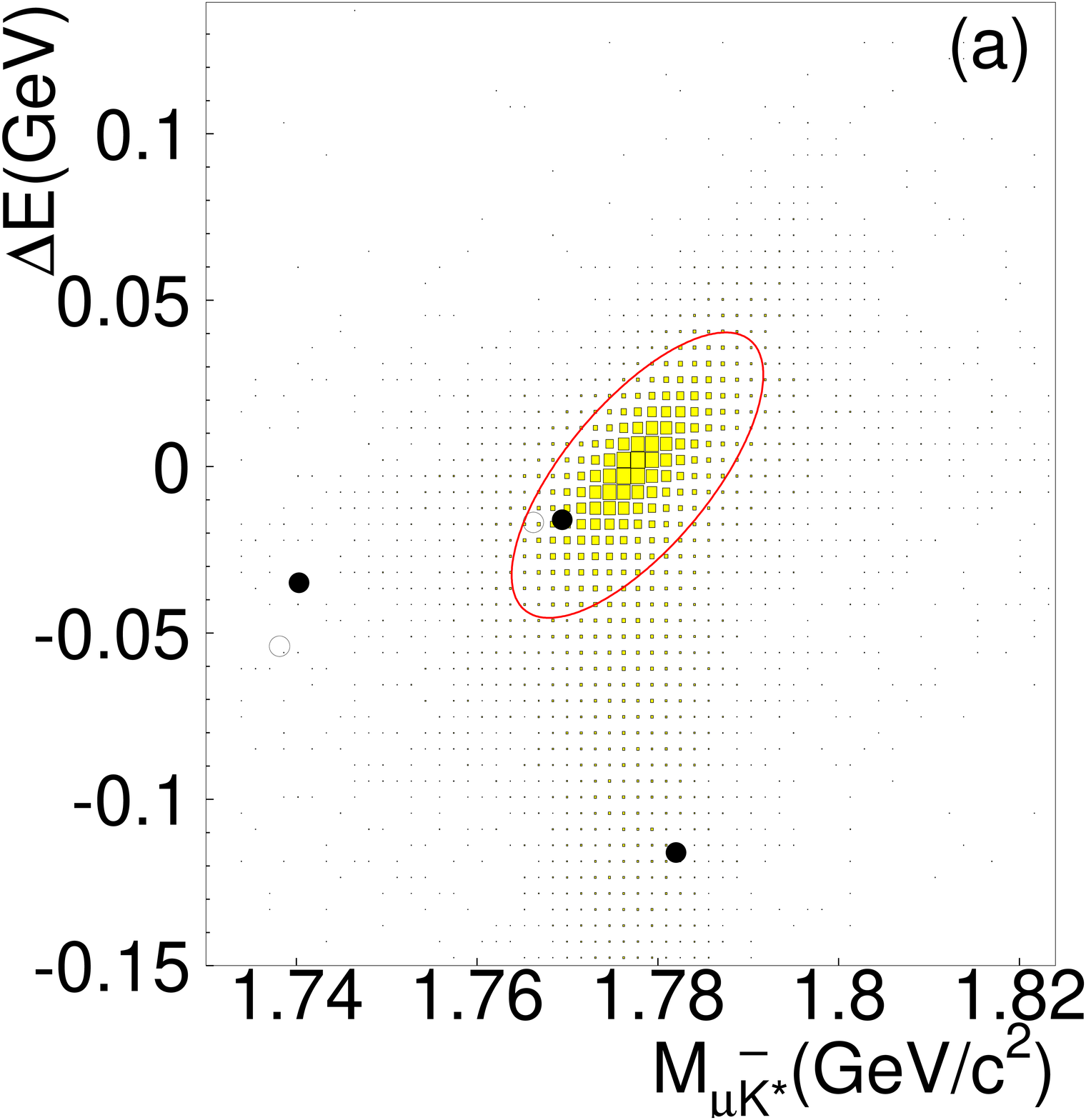}
\includegraphics[width=0.23\textwidth]{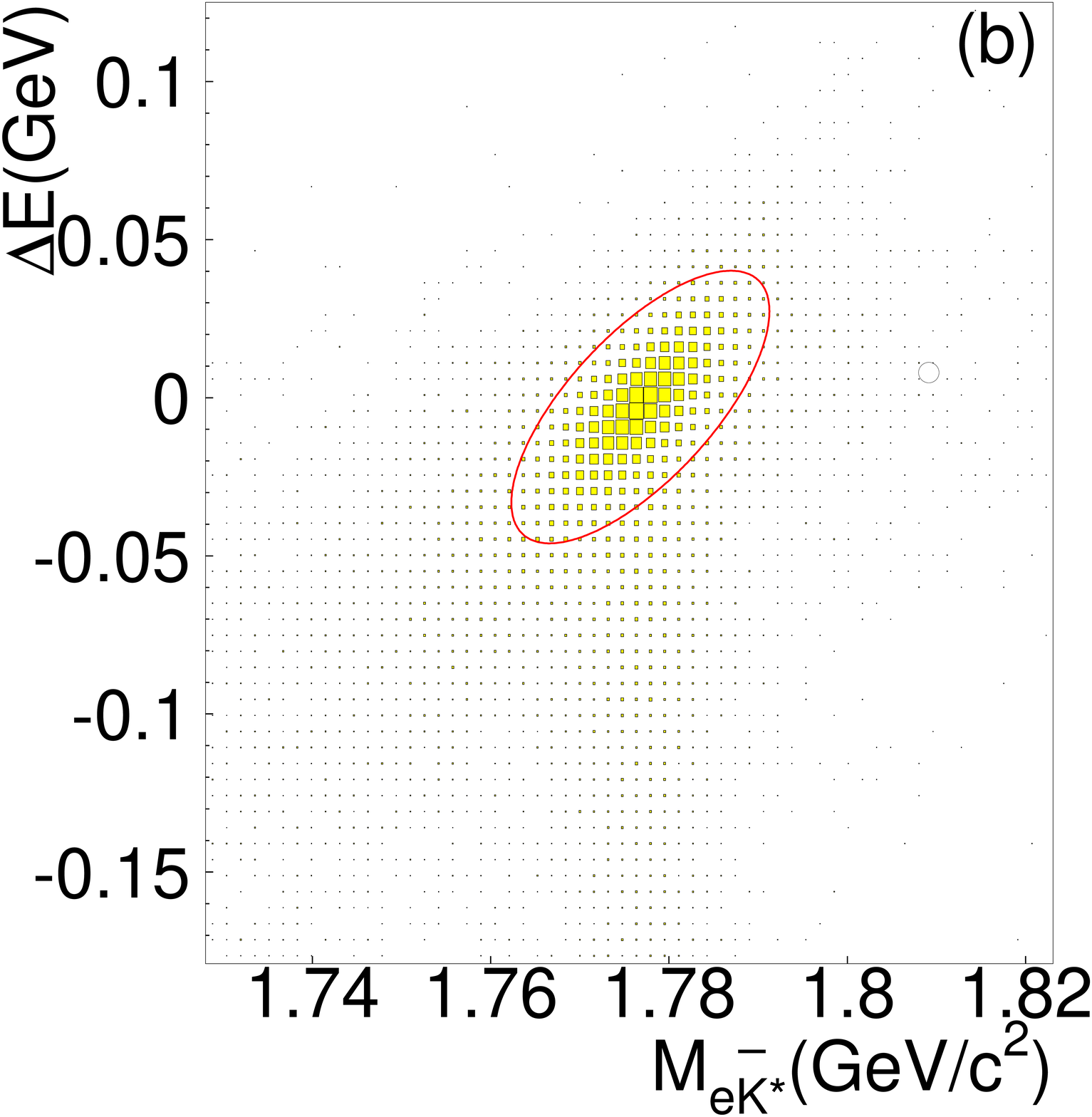}
\caption{$\Delta E$ vs. $M_{\rm inv}$ distributions for 
(a) $\tau^- \to \mu^- \bar{K}^{*0}$ and (b) $\tau^- \to e^- \bar{K}^{*0}$,
after all selection criteria.
Dots are data, yellow boxes show the signal MC and open circles show
$\tau$-pair background MC. The elliptical area is the $3\sigma$ signal region.}
\label{fig:result4}
\end{figure}

After all selection requirements, a few events remain in the signal region,
as shown in Figs.~\ref{fig:result1}, \ref{fig:result2}, \ref{fig:result3}
and \ref{fig:result4}.
For the $\ell \phi$, $\ell K^{*0}$ and $\mu \bar{K}^{*0}$ modes,
the expected number of BG events in the 
signal region are estimated using the sideband data, assuming 
that the BG distribution is flat in the $\pm 10 \sigma$ box.
For the $\mu \omega$ modes, we estimate the BG contribution in the signal
region using the BG MC distribution normalized to the ratio of data and MC
in the sideband region, because a rather large number of BG events
remain, which are mainly from $\tau^- \to \pi^- \omega \nu_{\tau}$.
For the $\tau^- \to e^- \omega$ and $e^- \bar{K}^{*0}$ modes,
since no events remain in 
the $\pm 10\sigma$ box for data, the number of expected BG events in 
the signal region is zero.
We calculate the error from the number of remaining MC events in 
the $\pm 10\sigma$ box assuming a flat BG distribution.
The number of events in the $\pm 10\sigma$ box excluding the signal region and 
the number of the expected BG events are listed in Tables \ref{tbl:bg1} and \ref{tbl:bg2}.
The comparison between the data and MC shows reasonable agreement;
the BG is suppressed well by the event selection.

\begin{table}[htb]
\caption{Number of events in the sideband region.}
\label{tbl:bg1}
\begin{center}
\begin{tabular}{c|cc}\hline
  & \multicolumn{2}{c}{$N_{\rm ev}$ in the $\pm10\sigma$ box} \\ 
 Mode & \multicolumn{2}{c}{in the sideband region} \\ 
 & ~~~data~~~  & MC \\
\hline 
$\tau^-\to\mu^-\phi$   & 2 & $1.68\pm1.17$ \\
$\tau^-\to e^-\phi$    & 2 & 0              \\
$\tau^-\to\mu^-\omega$ & 8 & $8.97\pm1.72$  \\
$\tau^-\to e^-\omega$  & 0 & $1.07\pm0.62$ \\
$\tau^-\to\mu^- K^{*0}$ & 4 & $3.94\pm1.61$ \\
$\tau^-\to e^-  K^{*0}$ & 1 & $1.58\pm1.11$ \\
$\tau^-\to\mu^- \bar{K}^{*0}$ & 2 & $0.90\pm1.11$ \\
$\tau^-\to e^-  \bar{K}^{*0}$ & 0 & $0.06\pm0.06$ \\
\hline
\end{tabular}
\end{center} 
\end{table}%

\begin{table}[htb]
\caption{Number of expected BG events.}
\label{tbl:bg2}
\begin{center}
\begin{tabular}{c|c}\hline   
      & $N_{\rm ev}$ of expected BG \\ 
Mode  & in the signal region  \\ 
\hline 
$\tau^-\to\mu^-\phi$   &  $0.11\pm0.08$   \\
$\tau^-\to e^-\phi$    &  $0.11\pm0.08$   \\
$\tau^-\to\mu^-\omega$ &  $0.19\pm0.21$   \\
$\tau^-\to e^-\omega$  &  $0\pm0.07$      \\
$\tau^-\to\mu^- K^{*0}$ & $0.22\pm0.11$  \\
$\tau^-\to e^-  K^{*0}$ & $0.05\pm0.05$  \\
$\tau^-\to\mu^- \bar{K}^{*0}$ & $0.10\pm0.07$  \\
$\tau^-\to e^-  \bar{K}^{*0}$ & $0\pm0.02$  \\
\hline
\end{tabular}
\end{center} 
\end{table}%

From the remaining number of data events in the signal region and 
the number of expected BG events,
we evaluated the upper limit $s_{90}$ on the number of signal events
at 90\% CL with systematic uncertainties
included in the Feldman-Cousins method~\cite{FeldmanCousins}
using the POLE code~\cite{HighlandCousins}.
The main systematic uncertainties on the detection efficiency come from 
track reconstruction (1.0\% per track), electron identification (2.2\%),
muon identification (2.0\%), kaon identification (1.4\% for $\phi$ 
reconstruction and 1.1\% for $K^{*0}$),
$\pi^0$ reconstruction (4.0\%), statistics of the signal
MC (1.3\% for $\ell \phi$, 0.7\% for $\ell \omega$ and 0.6\% for $\ell K^{*0}$
and $\ell \bar{K}^{*0}$)
and uncertainties in the branching fractions for $\phi \to K^+ K^-$
and $\omega \to \pi^+ \pi^- \pi^0$ (1.2\% and 0.8\%).
The uncertainty in the number of $\tau$-pair events mainly comes from the
luminosity measurement (1.4\%).
 
Upper limits on the branching fractions ${\cal B}$ are calculated as
${\cal B} < \frac{s_{90}}{2 N_{\tau\tau} \epsilon}$,
where $N_{\tau\tau}=4.99 \times 10^8$, which we calculate using
cross section of 0.919 nb according to ~\cite{tautau},
is the total number of the $\tau$-pairs produced 
and $\epsilon$ is the signal efficiency including 
the branching fractions of 
$\phi \to K^+ K^-$, $\omega \to \pi^+ \pi^- \pi^0$ and
$K^{*0} \to K^+ \pi^-$~\cite{PDG}.
The resulting upper limits on the branching fractions are summarized in
Table~\ref{tbl:summary}.

\begin{table}[htb]
\caption{Summary of the number of observed events $N_{\rm obs}$,
detection efficiency $\epsilon$,
total systematic error $\Delta\epsilon/\epsilon$,
90\% CL upper limit of signal events $s_{90}$
and 90\% CL upper limit of branching fractions.}
\label{tbl:summary}
\begin{center}
\begin{tabular}{cccccc}
\hline
Mode & ~~$N_{\rm obs}$~~ & $\epsilon$ & $\Delta\epsilon/\epsilon$ & ~~$s_{90}$~~ & UL on BF \\
  &  & (\%) & (\%) &  & (90\% CL)\\
\hline
$\tau^-\to\mu^-\phi$   & 1 & 3.14 & 5.2 & 4.21 & $1.3\times 10^{-7}$ \\
$\tau^-\to e^- \phi$   & 0 & 3.10 & 5.3 & 2.34 & $7.6\times 10^{-8}$ \\
$\tau^-\to\mu^-\omega$ & 0 & 2.51 & 6.3 & 2.25 & $9.0\times 10^{-8}$ \\
$\tau^-\to e^- \omega$ & 1 & 2.46 & 6.3 & 4.34 & $1.8\times 10^{-7}$ \\
$\tau^-\to\mu^- K^{*0}$& 0 & 3.71 & 4.8 & 2.24 & $6.1\times 10^{-8}$ \\
$\tau^-\to e^- K^{*0}$ & 0 & 3.04 & 4.9 & 2.42 & $8.0\times 10^{-8}$ \\
$\tau^-\to\mu^- \bar{K}^{*0}$ & 1 & 4.02 & 4.8 & 4.23 & $1.1\times 10^{-7}$ \\
$\tau^-\to e^- \bar{K}^{*0}$  & 0 & 3.21 & 4.9 & 2.45 & $7.7\times 10^{-8}$ \\
\hline
\end{tabular}
\end{center}
\end{table} 

\section{Summary}

We have searched for LFV decays $\tau^- \to \ell^- \phi$, $\ell^- \omega$,
$\ell^- K^{*0}$ and $\ell^- \bar{K}^{*0}$
using a 543~fb$^{-1}$ data sample from the Belle experiment.
No evidence for a signal is observed and upper limits on the 
branching fractions
are set in the range $(0.6 - 1.8) \times 10^{-7}$
at the 90\% confidence level.
This analysis is the first search for $\tau \to \ell \omega$ modes.
The results for the $\tau^- \to \ell^- \phi$,
$\ell^- K^{*0}$ and $\ell^- \bar{K}^{*0}$ modes are $3.6 - 9.6$ times more
restrictive than our previous results obtained using 158~fb$^{-1}$ of data.
The sensitivity improvement includes a factor of 3.4 in data
statistics and an optimized analysis
with higher efficiency and much improved BG suppression. 
The improved upper limits
can be used to constrain the parameter spaces of various scenarios beyond 
the SM.

\section{Acknowledgements}

We thank the KEKB group for the excellent operation of the
accelerator, the KEK cryogenics group for the efficient
operation of the solenoid, and the KEK computer group and
the National Institute of Informatics for valuable computing
and Super-SINET network support. We acknowledge support from
the Ministry of Education, Culture, Sports, Science, and
Technology of Japan and the Japan Society for the Promotion
of Science; the Australian Research Council and the
Australian Department of Education, Science and Training;
the National Science Foundation of China and the Knowledge
Innovation Program of the Chinese Academy of Sciences under
contract No.~10575109 and IHEP-U-503; the Department of
Science and Technology of India; 
the BK21 program of the Ministry of Education of Korea, 
the CHEP SRC program and Basic Research program 
(grant No.~R01-2005-000-10089-0) of the Korea Science and
Engineering Foundation, and the Pure Basic Research Group 
program of the Korea Research Foundation; 
the Polish State Committee for Scientific Research; 
the Ministry of Education and Science of the Russian
Federation and the Russian Federal Agency for Atomic Energy;
the Slovenian Research Agency;  the Swiss
National Science Foundation; the National Science Council
and the Ministry of Education of Taiwan; and the U.S.\
Department of Energy.


\begin{thebibliography}{99}
\bibitem{CC} Throughout this paper, 
the inclusion of the charge-conjugate decay modes is implied
unless otherwise stated.
\end{thebibliography}
\end{document}